# Carbon nanofiber-filled conductive silicone elastomers as soft, dry bioelectronic interfaces

Geoffrey A. Slipher, W. David Hairston, and Randy A. Mrozek
U.S. Army Research Laboratory, Aberdeen Proving Ground, MD

*Abstract*— Soft and pliable conductive polymer composites hold promise for application as bioelectronic interfaces such as for electroencephalography (EEG). In clinical, laboratory, and real-world EEG there is a desire for dry, soft, and comfortable interfaces to the scalp that are capable of relaying the µV-level scalp potentials to signal processing electronics. A key challenge is that most material approaches are sensitive to deformation-induced shifts in electrical impedance associated with decreased signal-to-noise ratio. This is a particular concern in real-world environments where human motion is present. The entire set of brain information outside of tightly controlled laboratory or clinical settings are currently unobtainable due to this challenge. Here we explore the performance of an elastomeric material solution purposefully designed for dry, soft, comfortable scalp contact electrodes for EEG that is specifically targeted to have flat electrical impedance response to deformation to enable utilization in real world environments. A conductive carbon nanofiber filled polydimethylsiloxane (CNF-PDMS) elastomer was evaluated at three fill ratios (3, 4 and 7 volume percent). Electromechanical testing data is presented showing the influence of large compressive deformations on electrical impedance as well as the impact of filler loading on the elastomer stiffness. To evaluate usability for EEG, pre-recorded human EEG signals were replayed through the contact electrodes subjected to quasi-static compressive strains between zero and 35%. These tests show that conductive filler ratios well above the electrical percolation threshold are desirable in order to maximize signal-to-noise ratio and signal correlation with an ideal baseline. Increasing fill ratios yield increasingly flat electrical impedance response to large applied compressive deformations with a trade in increased material stiffness, and with nominal electrical impedance tunable over greater than 4 orders of magnitude.

## I. Introduction

Electroencephalography (EEG) measures electrical voltage potentials at the scalp, which are created by pools of electrical currents from brain activity emanating outward. Potentials are traditionally measured using electrodes connected to the skin surface with conductive gels or pastes [1]. Although most commonly used for medical diagnosis and monitoring, EEG has also become a valuable tool for basic research applications [2]. In recent years there has been an increased interest in using EEG for measuring brain activity in real-world, active environments outside of traditional clinical scenarios [3]–[5] not only for basic research but also for use in brain-computer interaction (BCI) systems [6], [7].

State-of-the-art scalp contact sensors used in clinical and research EEG are hydrogel based, involve substantial setup time, require specialized training and experience, and experience degraded performance in a matter of hours. The sum total effect of all these factors leads to a burdensome task [8] to obtain high fidelity data under varying environmental conditions and over long periods of time. As a result, significant research interest has focused on "dry" sensors to contact the scalp and enable the collection of event related potentials for EEG in real-world interactive scenarios [2], [9]–[11]. While traditional gel electrodes use a layer of gel to make an electrolytic bridge with the scalp, dry sensors instead rely solely on a direct electrical connection, creating a strong dependence on the mechanical contact.

Several designs and approaches have recently been proposed, including conductive fabrics wrapped around foam [12], [13], spring-loaded pins [14], [15], plastics with conductive coatings [16] and firm carbon sensors [17], [18]. While reasonably effective, these all (except for fabric sensors) suffer from concerns regarding long-term comfort or safety. While fabric sensors can provide long-term comfort and safety, they are only viable for non-hairy sites which limits the measurement locations on the head.

We propose that a better long-term solution would be soft, deformable conducting elastomers. While a number of conductive elastomer substrates are available commercially, a common challenge is a large shift in electrical impedance when the materials are subjected to even small deformations [19] that would be anticipated during normal human movement. The large shift in electrical impedance can lead to a loss in the signal transmission or an artefact in the transmitted signal that can be even more troublesome as it would require additional data processing to separate the real and distorted portions of the EEG data.

We present a mixture of electrically conductive carbon nanofibers suspended within a silicone elastomer matrix as a viable and industrially scalable dry contact electrode approach. The carbon nanofibers were incorporated into the polydimethylsiloxane at three different concentrations, all of which are above the electrical percolation threshold. It is anticipated that as the concentration of the carbon nanofibers is increased, the number of conductive pathways increases along with a reduction in electrical impedance magnitude. The increased number of conductive pathways should allow for the transmission of the EEG signal even when some of the pathways are severed during deformation [20]. This is particularly relevant because dry EEG electrodes generally must be slightly compressed in order to maintain contact with the scalp, yet may be susceptible to some degree of flexion during active movements. As a result, it is important to investigate the relationship between mechanical deformation and electrical responses as a function of material formulation.

Additionally, given that some signal distortion is inevitable with compressive strain, it is also important to assess how this distortion will affect the relationship to the types of signals expected for the target use domain (e.g., real-world EEG) to ensure viability for this application prior to the expense and burden of human-subject testing. Typically this confirmation involves using the sensors on human subjects while performing some task in order to elicit a response to be observed. However, because the actual signals being elicited are unknown (e.g. there is no ground-truth comparator) and inconsistent (inherent non-stationarity of brain activity) this creates substantial sample variance that can be difficult to separate from the performance of the material itself. We propose an alternative approach involving reconstructed true EEG directly through the system components.

Here, we explore the plausibility of carbon nanofibers within a silicone elastomer matrix as a suitable substrate for EEG electrodes based on their response to compressive strain. Specifically, we describe the electrical performance in relation to compressive strain, and show how this relates directly to the clarity of recorded EEG when using conventional data acquisition measures. This latter test is performed using all components of a real system except the actual human subjects in order to ensure a true known signal and limit unaccountable signal variance.

## II. METHODS

### A. Materials Processing

Our efforts to produce soft conductive elastomers for scalp contact electrodes have focused on the use of materials with known biocompatibility. Specifically, the polymer component is composed of polydimethylsiloxane (PDMS), an example of a group of materials commonly referred to as "silicones" that have a flexible silicone-oxygen backbone chemistry. Carbon nanofibers (Pyrograf Products Inc.; PR-24-XT-HHT) are mixed in with the silicone elastomer in sufficient concentrations to promote an electrically percolating network throughout the otherwise insulating material. In this paper we consider a resultant nano-composite material for which the electrical impedance performance can be tailored over more than 4 orders of magnitude by controlling the volume fraction of carbon nanofiber used.

The addition of filler materials to an uncured elastomer has the effect of increasing the viscosity of the mixture. In order to overcome processing difficulties arising from a mixture that was too high viscosity, a low molecular weight nonreactive PDMS oil was added to the electrode formulations at 50 vol. % relative to the reactive precursors. The addition of the PDMS oil decreases the viscosity of the mixture, thereby enabling processing at higher carbon nanofiber concentrations up to approximately 8 vol. %. In addition, the concentration of nonreactive PDMS oil can be altered to tailor the mechanical stiffness of the resulting cured electrode , allowing for independent tuning of the electrical and mechanical response [21]. Once the synthesis and processing methods for the CNF-PDMS materials were refined, batches of electrodes were produced in 3 loadings (3, 4 and 7 vol. %).

### B. Electromechanical testing

Material samples were evaluated under quasi-static simple compressive loading. Specimens used for electromechanical testing were prepared using the procedure outlined above. Simple cylinders 8mm diameter by 16mm long were evaluated in all cases. Stress-strain was evaluated using a 20N load cell and micron accuracy extensometer simultaneously with electrical impedance characterization using a Keysight E5061B-LF network analyzer (frequency range limitation of 5Hz to 30MHz for electrical impedance). Two types of electromechanical characterizations were performed: 1) a cyclic loading/ unloading between zero and 35% compressive strain with simultaneous electrical impedance characterization at a single frequency (10Hz); and 2) a stepped loading profile in which a series of fixed strains were applied and strain was paused while electrical impedance scans were taken between 5Hz and 1.5kHz, the frequencies of interest for EEG. The single frequency characterizations yield smooth curves for strain dependent shifts in electrical impedance, whereas the stepped tests reveal changes in frequency dependent electrical impedance shifts as a function of different levels of applied strain. Both types of tests are required in order to understand the influence of mechanical deformation on shifts in scalp contact electrode electronic performance.

### C. Performance characterization for EEG

Material samples were placed in compressive strain as described above. Five samples of each loading fraction were evaluated. During each level of compression, a single channel of EEG data from a prior recorded 6-min session was reproduced using a NI USB-6356 and passed through the electrode. Prior to reproduction, data were filtered 0.5-45 Hz to keep within a reasonable dynamic range for the DAC and attenuated using two 40dB attenuators in series to match a realistic approximate amplitude as is typically observed on the scalp (e.g. +/-50 µV). The reproduced data originate from scalp location P03 of a randomly chosen human subject performing a rapid serial visual presentation (RSVP) task [22]. In this task, images were presented at a rate of 2 Hz and included a randomized assortment of simulated scene images; the subjects pressed a button when they saw a scene including a target image. Reproduced data were recorded using an EEG system based on an ADS1298 chipset-based commercial EEG system (Mindo32) by directly connecting the recording system leads to the sample (with the NI USB-6356 output on the other side). In this manner, the general recording setup was roughly similar to what would be used in a typical setting but without variable or noise-inducing components, such as the human subject or conductive gels.

In order to assess signal efficacy, we examined the correlation between the recorded signal using sample electrodes and a baseline pass-through recording (no electrode) of the native (input) EEG at each level of compressive strain. This statistic was chosen as a global measure of signal reproduction because it is agnostic to inevitable differences in amplitude due to miscalibration between ADCs or amplifiers, to long-term low-frequency drift, or to phase-shifts from minor signal asynchronies. A direct-connection (no electrode) pass-through in which the wire leads were directly shorted together was used as baseline instead of the original native input waveform because this also includes distortions and nonlinearities in the signal which are inherent to the acquisition system and method. Data files (baseline and each compression) were initially inverse band-pass filtered 0.5–45 Hz to remove large trends and line noise artifacts, and time-aligned to one another using cross-correlation. Files were then divided into 176 independent segments of 1,400 ms each. Pearson correlation coefficient was calculated between each recorded file and the baseline file for each segment. A Fisher Z' transformation was applied to the correlation coefficients in order to normalize the sampling distribution; these values were then averaged across samples at each level of compression to yield a mean and standard error in correlation for the recordings. We additionally included a set of direct-connection records to use as comparators for maximally ideal correlation scores using this method. Thus the final statistics denote how well, on average, the recorded signal tracks the original EEG data for each volume percentage of filler as the material is compressed.

III. RESULTS

*A. Electromechanical testing*

The mechanical impact of adding conductive filler is presented in Fig. 1. As the filler volume ratio is increased, two trends are observed. First, the stiffness (slope of stress-strain curve) is dramatically increased with increasing filler ratios. This has an impact on the resulting electrode comfort, with increasing stiffness associated with decreasing comfort. Second, increasing filler ratios are associated with an increase in mechanical hysteresis in the initial loading/unloading cycle and residual plastic deformation in the material as observed in the loading versus unloading paths for the material. This hysteresis and plastic deformation must be accounted for in any engineering analysis or design work. Provided the peak strain magnitude does not change, loading/unloading paths of subsequent cycles closely follow the unloading path of the initial cycle in the same manner as results on magnetic elastomers presented by Lazarus et al [23].

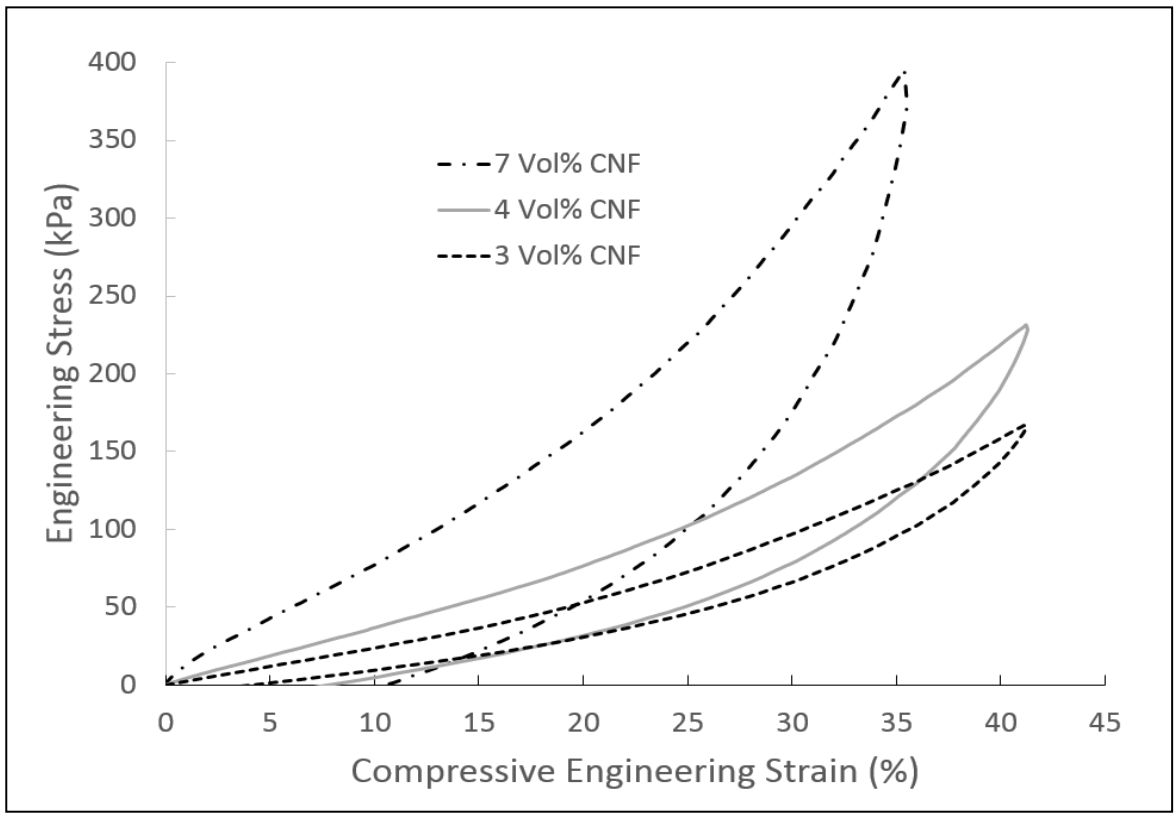

Figure 1. Compressive stress-strain curves for three different electrode filler loadings, 3, 4 and 7 vol % carbon nanofilber in PDMS

Frequency response curves were observed to be flat between 5 Hz and 1.5kHz. The dominant effect is, therefore, strain induced shifts in electrical impedance (Fig. 2). We note that the strain induced effect is most prominent when conductive filler ratios are utilized such that operation occurs closer to the electrical percolation threshold (e.g. 3 volume percent case). Strain induced shifts in phase angle were in all cases less than +/- 5 degrees, approximately centered around the zero phase angle axis.

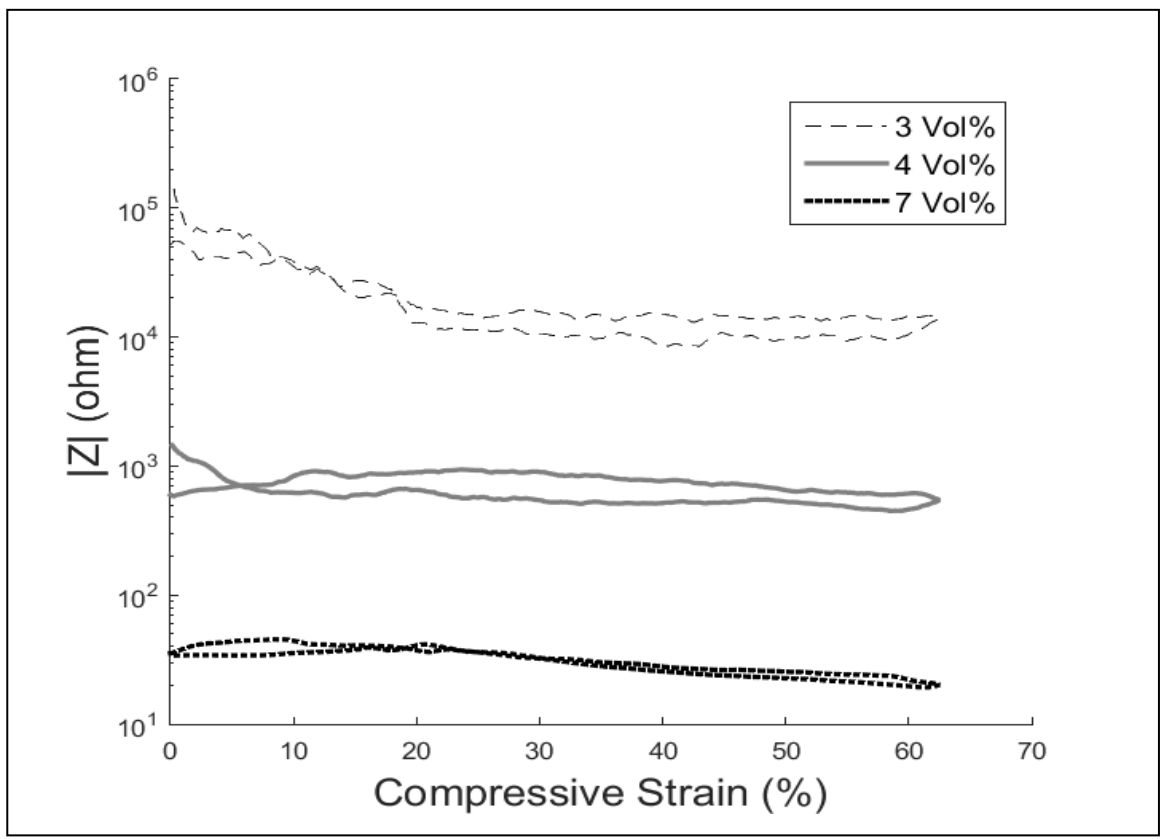

Figure 2. Single frequency (10Hz) electrical impedance performance over a single strain cycle (increasing/decreasing) for three different electrode filler loadings: 3, 4 and 7 volume percent carbon nanofiber in PDMS.

Not surprisingly there was a notable difference in overall electrical impedance based on on filler percent loading, such that the higher concentration led to more than a 100x decrease in impedance (Fig. 2). Note also that for 3% loading, impedance dercreased dramatically (up to 10x) with electrode compression, with maximal conductance plateauing above 15% compressive strain.

*B. Performance characterization for EEG*

Correlation to the baseline (direct connection) signal depended highly on both the filler percent loading (3%-7%) and the degree of compressive strain (Fig. 3). Specifically, with lower (3%) filler (dashed lines), the initial level of correlation was marginal (0.679, z' of 0.86), but increased steadily to plateau above 20% compressive strain and a mean correlation of 0.952 (z' of 1.89). However, it never achieved comparable performance to the baseline mean (0.990 correlation or 2.71 z'), In contrast, higher (7%) filler loading showed performance nearly matching baseline with no observable effect of compressive strain (mean 0.986 correlation and 2.52 z' across the range).

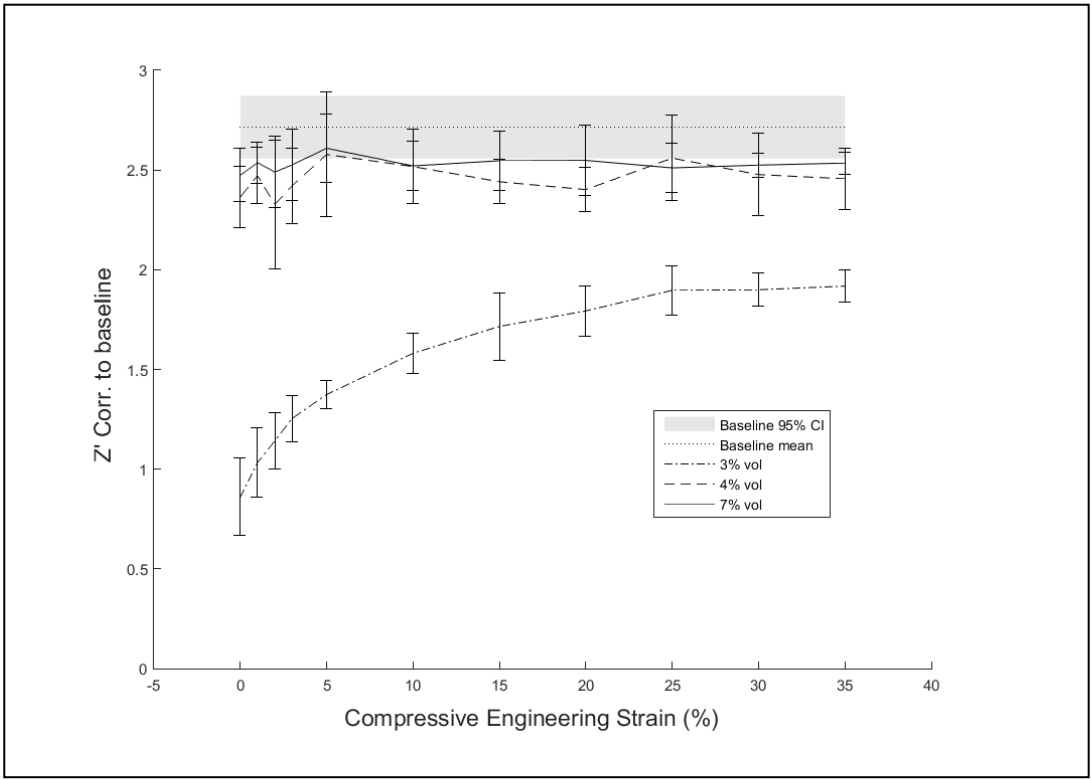

Figure 3. Correlation (Z' transformed R) between recorded signals and a baseline direct connection pass-through record for each filler loadings (3, 4, and 7 vol%) with increasing comrpessive strain. Horizontal dashed line represents ideal expected performance based on correlation to other pass-through records.

## IV. Discussion

Here, we have explored the potential for using PDMS silicone elastomers filled with electrically conductive carbon nanofibers as the basis for EEG electrodes. Our results suggest general viability for this approach, and substantial value for further pursuit. For lower nanofiber loadings, there is a clear dependence on compressive strain to increase the general conductance to a plateau level. Although conductive performance is less than ideal under no strain, the overall impedance level still falls well below what is commonly observed for the skin-electrode barriers for dry EEG approaches, which tend to be in the giga-Ohm range [9]. Even at this level, we still found reasonable correlations to the native signal, with variance across the signal no worse than for lower-impedance conditions.

As compressive force is added, signal characteristics only improve, hitting a maxima with low pressures and strains (between 50 and 100 kPa, and 15%, respectively). This flat plateau effect suggests general robustness to a variety of uses where strain may not be consistent. Meanwhile, the overall dependence on compression is completely ameliorated by using a filler loading significantly above the electrical percolation threshold, in this case both 4 and 7 volume percent. Sufficiently high loading resulted in a very low overall impedance with high consistency across strain.

The major drawback of using a higher filler loading is a substantially more firm electrode. This will in turn dramatically affect comfort and the potential long-term usability of the materials as electrodes, and thus must be considered carefully. We thus recommend future work to focus on the range between these values and the potential balance point between ideal electrical and tolerable mechanical performance, since each place a measureable bound on usability.

In this work, for validation with EEG we have specifically focused on using re-created, real pre-recorded EEG data and standard commercial EEG system components, but in the absence of human subjects. This was done to remove uncontrollable human-related variability and retain the ability to compare against a known-quantity signal. We have focused on the correlation between recorded and the most ideal signal possible with no intervening electrode present as a metric for overall signal reproduction fidelity. In normal human-subjects testing, such a comparison would not be feasible due to the inability to reproduce an identical data series over time in the exact same scalp location. The observed variance (shown in error bars) thus arises from a combination of minor differences in the electrodes themselves and small differences in recorded noise during data acquisition. Our observation was that the trends for changes in correlation coefficients over time were highly consistent across repetitions and compression samples; that is, epochs with low correlation were always low, and those that were higher were always high. Thus we feel reasonably confident that the majority of this variance is related to the nature of the data collection and not an effect of the materials themselves. Rather, as would be predicted in the case of higher electrical impedance, we saw that the cases of overall lower correlation (e.g. 3% filler and no compression) were primarily driven by an overall higher noise level and lower signal-to-noise ratio.

Other metrics of signal reproduction may be more appropriate, but a common challenge is that there is little agreement in the field regarding how to best validate the efficacy of EEG systems. Some groups rely on either basic non-comparative values such as SNR from RMS noise ratios or the performance of BCI classifiers for various equipment [9], [12], [14], [24]. It remains unclear, however, exactly how these kinds of classification actually relate to the overall "usability" of an EEG system [25]. Since this work was focused on an initial exploration of viability, we have limited to simple correlation. Future work, however, will explore the use of more advanced classifiers and a metric of usability.

## V. Conclusion

A material solution has been presented for dry, soft, and comfortable bioelectronics interfaces, and was specifically evaluated as scalp contact electrodes for application to EEG collection in real-world conditions. A conductive carbon nanofiber material was added to a silicone elastomer rendering the resulting composite material conductive. The ratio of the filler material must be well above the electrical percolation threshold in order to avoid deformation induced shifts in electrode electrical impedance that can negatively impact EEG signal collection performance during use. At the same time, the ratio of the filler material must be minimized in order to minimize the impact on stiffness and maintain a soft and comfortable material solution. Thus, an optimal material solution would be somewhere in between the 3, 4, and 7 volume percent formulations discussed in this paper.

## VI. Acknowledgements

This work was supported by U.S. Army Research Laboratory Director's Research Innovation grant number DRI-FY13-VTD-009.

## References

[1] T. C. Ferree, P. Luu, G. S. Russell, and D. M. Tucker, "Scalp electrode impedance, infection risk, and EEG data quality," *Clin. Neurophysiol.*, vol. 112, no. 3, pp. 536–544, Mar. 2001.
[2] K. McDowell, Chin-Teng Lin, K. S. Oie, Tzyy-Ping Jung, S. Gordon, K. W. Whitaker, Shih-Yu Li, Shao-Wei Lu, and W. D. Hairston, "Real-World Neuroimaging Technologies," *IEEE Access*, vol. 1, pp. 131–149, 2013.
[3] J.-P. Lachaux, N. Axmacher, F. Mormann, E. Halgren, and N. E. Crone, "High-frequency neural activity and human cognition: Past, present and possible future of intracranial EEG research," *Prog. Neurobiol.*, vol. 98, no. 3, pp. 279–301, Sep. 2012.
[4] S. K. L. Lal and A. Craig, "Driver fatigue: Electroencephalography and psychological assessment," *Psychophysiology*, vol. 39, no. 3, pp. 313–321, May 2002.
[5] A. Gevins, H. Leong, R. Du, M. E. Smith, J. Le, D. DuRousseau, J. Zhang, and J. Libove, "Towards measurement of brain function in operational environments," *Biol. Psychol.*, vol. 40, no. 1–2, pp. 169–186, May 1995.
[6] M. M. Moore, "Real-world applications for brain~computer interface technology," *IEEE Trans. Neural Syst. Rehabil. Eng.*, vol. 11, no. 2, pp. 162–165, Jun. 2003.
[7] B. J. Lance, S. E. Kerick, A. J. Ries, K. S. Oie, and K. McDowell, "Brain–Computer Interface Technologies in the Coming Decades," *Proc. IEEE*, vol. 100, no. Special Centennial Issue, pp. 1585–1599, May 2012.
[8] W. David Hairston, K. W. Whitaker, A. J. Ries, J. M. Vettel, J. Cortney Bradford, S. E. Kerick, and K. McDowell, "Usability of four commercially-oriented EEG systems," *J. Neural Eng.*, vol. 11, no. 4, p. 46018, Aug. 2014.
[9] Y. M. Chi, Yu-Te Wang, Yijun Wang, C. Maier, Tzyy-Ping Jung, and G. Cauwenberghs, "Dry and Noncontact EEG Sensors for Mobile Brain–Computer Interfaces," *IEEE Trans. Neural Syst. Rehabil. Eng.*, vol. 20, no. 2, pp. 228–235, Mar. 2012.
[10] L.-D. Liao, C.-Y. Chen, I-J. Wang, S.-F. Chen, S.-Y. Li, B.-W. Chen, J.-Y. Chang, and C.-T. Lin, "Gaming control using a wearable and wireless EEG-based brain-computer interface device with novel dry foam-based sensors," *J. NeuroEngineering Rehabil.*, vol. 9, no. 1, p. 5, 2012.
[11] M. Lopez-Gordo, D. Morillo, and F. Valle, "Dry EEG Electrodes," *Sensors*, vol. 14, no. 7, pp. 12847–12870, Jul. 2014.
[12] C.-T. Lin, L.-D. Liao, Y.-H. Liu, I-J. Wang, B.-S. Lin, and J.-Y. Chang, "Novel dry polymer foam electrodes for long-term EEG measurement," *IEEE Trans. Biomed. Eng.*, vol. 58, no. 5, pp. 1200–1207, May 2011.
[13] J. Löfhede, F. Seoane, and M. Thordstein, "Textile Electrodes for EEG Recording — A Pilot Study," *Sensors*, vol. 12, no. 12, pp. 16907–16919, Dec. 2012.
[14] L.-D. Liao, I-J. Wang, S.-F. Chen, J.-Y. Chang, and C.-T. Lin, "Design, fabrication and experimental validation of a novel dry-contact sensor for measuring electroencephalography signals without skin preparation," *Sensors*, vol. 11, no. 6, pp. 5819–5834, 2011.
[15] P. Fiedler, P. Pedrosa, S. Griebel, C. Fonseca, F. Vaz, E. Supriyanto, F. Zanow, and J. Haueisen, "Novel Multipin Electrode Cap System for Dry Electroencephalography," *Brain Topogr.*, vol. 28, no. 5, pp. 647–656, Sep. 2015.
[16] T. R. Mullen, C. A. E. Kothe, Y. M. Chi, A. Ojeda, T. Kerth, S. Makeig, T.-P. Jung, and G. Cauwenberghs, "Real-Time Neuroimaging and Cognitive Monitoring Using Wearable Dry EEG," *IEEE Trans. Biomed. Eng.*, vol. 62, no. 11, pp. 2553–2567, Nov. 2015.
[17] G. Ruffini, S. Dunne, E. Farres, P. C. P. Watts, E. Mendoza, S. R. P. Silva, C. Grau, J. Marco-Pallares, L. Fuentemilla, and B. Vandecasteele, "ENOBIO - First Tests of a Dry Electrophysiology Electrode using Carbon Nanotubes," 2006, pp. 1826–1829.
[18] B. Mahar, C. Laslau, R. Yip, and Y. Sun, "Development of Carbon Nanotube-Based Sensors—A Review," *IEEE Sens. J.*, vol. 7, no. 2, pp. 266–284, Feb. 2007.
[19] N. Chau, G. A. Slipher, B. M. O'Brien, R. A. Mrozek, and I. A. Anderson, "A solid-state dielectric elastomer switch for soft logic," *Appl. Phys. Lett.*, vol. 108, no. 10, p. 103506, Mar. 2016.
[20] Geoffrey Slipher, Randy Mrozek, W. David Hairston, Joseph Conroy, Wosen Wolde, and William Nothwang, "Stretchable Conductive Elastomers for Soldier Biosensing Applications: Final Report," ARL-TR-7615, U.S. Army Research Laboratory, 2016.
[21] R. A. Mrozek, P. J. Cole, K. J. Otim, K. R. Shull, and J. L. Lenhart, "Influence of solvent size on the mechanical properties and rheology of polydimethylsiloxane-based polymeric gels," *Polymer*, vol. 52, no. 15, pp. 3422–3430, Jul. 2011.
[22] A. R. Marathe, A. J. Ries, V. J. Lawhern, B. J. Lance, J. Touryan, K. McDowell, and H. Cecotti, "The effect of target and non-target similarity on neural classification performance: a boost from confidence," *Front. Neurosci.*, vol. 9, p. 270, 2015.
[23] N. Lazarus, C. D. Meyer, S. S. Bedair, G. A. Slipher, and I. M. Kierzewski, "Magnetic Elastomers for Stretchable Inductors," *ACS Appl. Mater. Interfaces*, vol. 7, no. 19, pp. 10080–10084, May 2015.

[24] A. J. Ries, J. Touryan, J. Vettel, K. McDowell, and W. D. Hairston, "A Comparison of Electroencephalography Signals Acquired from Conventional and Mobile Systems," *J. Neurosci. Neuroengineering*, vol. 3, no. 1, pp. 10–20, Feb. 2014.

[25] W. D. Hairston and V. Lawhern, "How Low Can You Go? Empirically Assessing Minimum Usable DAQ Performance for Highly Fieldable EEG Systems," in *Foundations of Augmented Cognition*, vol. 9183, D. D. Schmorrow and C. M. Fidopiastis, Eds. Cham: Springer International Publishing, 2015, pp. 221–231.